\documentclass[prd,twocolumn,showpacs,letterpaper,nofootinbib]{revtex4}

\usepackage{graphicx}

\newcommand{\ApJL}{Astrophys. J. Lett.}
\newcommand{\ApJ}{Astrophys. J.}
\newcommand{\PRL}{Phys. Rev. Lett.}
\newcommand{\PRD}{Phys. Rev. D}

\newcommand{\spin}[1]{\,{}_{#1}^{\vphantom{m}}}   
\renewcommand{\l}{{\bf l}}
\newcommand{\la}{{{\bf l}_1}}

\newcommand{\tot}{{\rm t}}

\newcommand{\cmb}{\Theta}

\newcommand{\vecl}{{\bf l}}

\newcommand{\intl}[1]{\int \frac{d^2 {\bf l}_#1}{(2\pi)^2}}

\newcommand{\bfr}{{\mathbf{r}}}
\newcommand{\bfH}{{\mathbf{H}}}
\newcommand{\bfk}{{\mathbf{k}}}

\newlength{\tskip}
\newlength{\colwidth}

\newcommand{\beq}{\begin{equation}}
\newcommand{\eeq}{\end{equation}}
\newcommand{\beqa}{\begin{eqnarray}}
\newcommand{\eeqa}{\end{eqnarray}}

\newcommand{\bn}{{\hat{\bf n}}}
\newcommand{\bm}{\hat{\bf m}}

\newcommand{\rad}{r}    

\newcommand{\len}{\phi}

\begin{document}

\title{Cosmic Shear of the Microwave Background: The Curl Diagnostic}
\author{Asantha Cooray$^{1,2}$, Marc Kamionkowski$^2$, and
Robert R. Caldwell$^3$}
\affiliation{
$^1$Department of Physics and Astronomy, 4129  Frederick Reines Hall, 
University of California,   Irvine, CA 92697\\
$^2$Mail Code 130-33, California Institute of 
Technology, Pasadena, California 91125\\
$^3$Department of Physics and Astronomy, Dartmouth College, 6127 Wilder 
Laboratory, Hanover, NH 03755}

\date{\today}

\begin{abstract} 
Weak-lensing distortions of the cosmic-microwave-background (CMB) temperature
and polarization patterns can reveal important clues to the
intervening large-scale structure. The effect of lensing is to deflect the
primary temperature and polarization signal to slightly different locations on
the sky. Deflections due to density fluctuations, gradient-type for the
gradient of the projected gravitational potential, give a direct measure of the
mass distribution. Curl-type deflections can be induced by, for example, a
primordial background of gravitational waves from inflation or by second-order
effects related to lensing by density perturbations. Whereas gradient-type
deflections are expected to dominate, we show that curl-type deflections can
provide a useful test of systematics and serve to indicate the presence of
confusing secondary and foreground non-Gaussian signals.

\end{abstract}


\maketitle

\section{Introduction}

The theory of weak gravitational lensing (``cosmic shear'') of
the cosmic microwave background (CMB) by large-scale
mass inhomogeneities has now been studied extensively
\cite{lensing,Hu00,Hu01,HuOka02,KesCooKam03,HirSel03}. The
distortions produce a unique
non-Gaussian signal in the CMB temperature and polarization patterns. The
distinction with the primordial Gaussian pattern can be exploited to
determine various properties of the lensing sources. Since inhomogeneities in
the intervening mass field are expected to provide the dominant source of
lensing, likelihood methods and quadratic estimators have been
developed to map the projected mass distribution. Such
reconstruction will be important not only to study the matter
power spectrum, but also to reconstruct the primordial
polarization signal from inflationary gravitational waves
\cite{KesCooKam02}.

To linear order in the density-perturbation amplitude, lensing
by mass fluctuations results in a deflection angle that can be
written as the gradient of a projected gravitational potential
(i.e., the deflection angle is a longitudinal vector field in
the plane of the sky).
Algorithms to reconstruct the mass distribution hence measure
only this longitudinal component of the deflection angle.  A
deflection angle that can be written as a curl---a gradient-free
or transverse-vector field---can be produced through lensing by gravitational
waves or through lensing by mass fluctuations to second order in the
density-perturbation amplitude.  Since the curl-type deflection
is expected to be significantly weaker than the gradient-type (as
discussed further below), measurement of the curl power spectrum
has been used as a test of systematic artifacts in the
cosmic-shear maps that have been produced with measurements of
shape distortions to high-redshift galaxies
\cite{cosmicshearrefs,cosmicsheardetections}.  For galaxy
cosmic-shear maps, this curl component is measured by simply
rotating each galaxy image by $45^\circ$ \cite{stebbins,consistency}.

For cosmic-shear of the CMB, one cannot simply rotate the
temperature pattern at each point on the sky.  However, there is
indeed a method to reconstruct the curl component of the
deflection angle that is directly analogous to that for
reconstructing the gradient component \cite{HirSel03}.  In this
paper, we show that the cosmological
curl signals are expected to be small, and that measurement of
the cosmic-shear curl component can thus be used as a diagnostic for
systematic artifacts, unsubtracted foregrounds, and/or
primordial non-Gaussianity.

This paper is organized as follows. In Section~\ref{sec2} we introduce our
formalism and present expressions for the weak-lensing
corrections to the CMB power spectra from both gradient and curl
modes of the deflection field.  In
Section~\ref{sec4} we discuss quadratic estimators for the
deflection field for both the gradient and curl components, and
we discuss the orthogonality of these estimators or filters.
Section~\ref{sec3} discusses cosmological sources for a curl
component, first gravitational waves and then second-order
density perturbations.  Section~\ref{sec:results} presents results
of our calculations.  Section~\ref{sec:diagnostic} discusses the
use of a curl reconstruction as a diagnostic for primordial
non-Gaussianity, unsubtracted foregrounds, or systematic
artifacts.  A few concluding remarks about the applications of
our work are given in Section~\ref{sec5}.

\section{EFFECT ON CMB POWER SPECTRA}
\label{sec2}

The effect of weak lensing on CMB anisotropies is a non-linear remapping of
temperature and polarization fluctuations. In the case of temperature
anisotropies on the sphere, this remapping can be expressed as
\begin{equation}
\tilde \cmb(\bn)  =   \cmb[\bn + \nabla \len(\bn) + \nabla \times \Omega(\bn)]
\label{eqn:deflect}  \, ,
\end{equation}
where $\tilde\cmb$ is the observed temperature pattern,
$\bn=(\theta,\phi)$ is the position on the sky, and $\cmb$ is the
original unlensed pattern.  Here, the gradient $\nabla \phi$ has
components $\partial_i \phi$ in the plane of the sky, and the
``curl'' $\nabla\times \Omega$ has components $\epsilon_{ij}
\partial_j \Omega$, where $\epsilon_{ij}$ is the antisymmetric
tensor.  The mapping involves the angular gradient of
the projected gravitational potential $\len$ due to density
perturbations, and the curl of some other function $\Omega$,
to be discussed further below.  The convergence $\kappa$ and the
image rotation $\omega$ that usually arise in gravitational
lensing of discrete sources are given by $\kappa(\bn) =
-\frac{1}{2} \nabla^2 \len(\bn)$ and $\omega(\bn) = -\frac{1}{2}
\nabla^2 \Omega(\bn)$, respectively \cite{CooHu02,Dodetal03}. In the
limit of weak deflection, the remapping in Eq.~(\ref{eqn:deflect})
can be expressed in Fourier space as
\begin{eqnarray}
     \tilde \cmb(\vecl)
     &=& \int d \bn\, \tilde \cmb(\bn) e^{-i \vecl \cdot \bn} \nonumber\\
     &=& \cmb(\vecl) - \int \frac{d^2 \vecl'}{(2\pi)^2} \cmb(\vecl')
     L^{\phi,\Omega}(\vecl,\vecl') \, ,
\label{eqn:thetal}
\end{eqnarray}
under the flat-sky approximation, where
\begin{eqnarray}
\label{eqn:lfactor}
     && L^\phi(\vecl,\vecl') \equiv \len(\vecl-\vecl') \,
     [(\vecl - \vecl') \cdot \vecl']\, , \\
\label{eqn:lfactoromega}
     && L^\Omega(\vecl,\vecl') \equiv \Omega(\vecl-\vecl')
     \, [(\vecl - \vecl') \times \vecl'].
\end{eqnarray}
Note that the curl component has a two-dimensional
cross-product, a ninety-degree rotation
between Fourier components which we denote as $\times$,
following Ref.~\cite{HirSel03}, whereas the gradient has a
dot-product.  Strictly speaking, there is an additional term
quadratic in $\phi$ and $\Omega$, respectively, that is required
to obtain the lowest-order cosmic-shear corrections to
the power spectrum; see, e.g., Eq.~(3) in Ref.~\cite{KesCooKam03},
and for $L^\Omega$ simply replace the dot products therein by
cross products.  For economy, we do not reproduce those
expressions here, but they are included in our numerical work.

The temperature-anisotropy power spectrum is
\begin{equation}
\widetilde C_l^\cmb = \left[1 - R\right] C_l^\cmb
        + \intl{1} C_{|\vecl - \vecl_1|}^\cmb C^{XX}_{l_1}
                [(\vecl - \vecl_1)\odot \vecl_1]^2  \, .
\label{eqn:ttflat}
\end{equation}
Here, $C^{XX}_{l}$ is the power spectrum of either lensing
potentials related to density fluctuations or the rotational
component, while $R$ is a multiplicative correction ${\cal
O}(\phi^2)$ that can be obtained from Eq.~(8) in
Ref.~\cite{KesCooKam03} (again replacing a dot product by a
cross product for $\Omega$).

In addition to temperature anisotropies, lensing also modifies
the polarization.  We follow the notation in Ref.~\cite{Hu00},
and then the remapping of the polarization under lensing is
\begin{equation}
     {}_{\pm} \tilde X(\bn)  =   {}_{\pm}X[\bn +
     \nabla\len(\bn) + \nabla \times \Omega(\bn) ]
\end{equation}
where $\spin{\pm} X = Q\pm i U$.   Since the Stokes parameters
are not rotationally invariant, we write them in terms of the rotational
invariants $E$ and $B$ \cite{KamKosSte97} which are defined in
Fourier space through ${}_\pm X(\l) = [E(\l) \mp i B(\l)] e^{\pm
2 i \varphi}$, where $\varphi$ is the phase angle of $\l$.
Then, the observed polarization is
\begin{equation}
     {}_\pm \widetilde X(\l) = {}_\pm X(\l)-\intl{1}
     {}_\pm X(\la)
      e^{\pm 2i (\varphi_{\vecl_1}- \varphi_{\vecl})}
     L^{\phi,\Omega}(\l,\l_1) \, .
\end{equation}
Following Ref.~\cite{Hu00}, the lensed polarization power spectra can now be
expressed in terms of $C_l^{XX}$ for $X=(\phi,\Omega)$  and the unlensed CMB
spectra as
\begin{widetext}
\begin{eqnarray}
\widetilde C_l^{EE} &=& \left[ 1 - R \right]   \,
                                C_l^{EE} 
        + \frac{1}{2} \intl{1} C_{|\vecl - \vecl_1|}^{XX}
[(\vecl - \vecl_1)\odot \vecl_1]^2 
 [C^{EE}_{l_1} \cos^2(2\varphi_{\vecl_1}) 
 + C^{BB}_{l_1} \sin^2(2\varphi_{\vecl_1})] \, ,\\
\widetilde C_l^{BB} &=& \left[ 1 - R \right]   \,
                                C_l^{BB} 
        + \frac{1}{2} \intl{1} C_{|\vecl - \vecl_1|}^{XX}
[(\vecl - \vecl_1)\odot \vecl_1]^2 
[C^{EE}_{l_1} \sin^2(2\varphi_{\vecl_1}) 
+ C^{BB}_{l_1} \cos^2(2\varphi_{\vecl_1})]  \, .
\end{eqnarray}
\end{widetext}
In the above, the operator $\odot$ is a dot product when
$X=\phi$ and a cross product when $X=\Omega$.  Note that
inclusion of the $R$ correction, which was neglected in
Ref.~\cite{HirSel03}, is required to obtain a correction that is
complete to lowest nonvanishing order in $\phi$ or $\Omega$.  We
will present numerical results for these power spectra in
Section~\ref{sec:results} after discussing the power spectra
$C_l^{\phi\phi}$ and $C_l^{\Omega\Omega}$ in Section~\ref{sec3}.


\section{Reconstruction of the Deflection Field}
\label{sec4}

So far we have discussed the corrections to the
temperature/polarization power spectra due to weak gravitational
lensing.  However, what is perhaps more interesting is that
lensing induces characteristic non-Gaussian signatures in the
temperature/polarization pattern.  Measurement of these
non-Gaussianities can be used to map the deflection-angle as a
function of position on the sky, and thus to infer the projected
potentials $\phi$ and $\Omega$.

We now extend the quadratic estimators that have been
proposed to reconstruct the gradient component of the deflection
field \cite{lensing,Hu00,Hu01,HuOka02} to the case of a curl component
\cite{HirSel03}.  In Fourier space, the quadratic estimator can
be written for $X=\phi$ or $X=\Omega$ as
\begin{equation}
     \hat{X}(\vecl) = \int \frac{d^2\vecl_1}{(2\pi)^2}
     W^{X}(\vecl, \vecl_1) \tilde \cmb(\vecl_1) \tilde
     \cmb(\vecl-\vecl_1) \, ,
     \label{eqn:quadest}
\end{equation}
where $W^X$ is a filter that acts on the CMB temperature field 
subject to the demands that $\langle \hat{\phi}(\vecl) \rangle
= \phi(\vecl)$ and  $\langle \hat{\Omega}(\vecl) \rangle =
\Omega(\vecl)$. The filters that optimize the signal-to-noise are
\begin{equation}
     W^X(\vecl, \vecl_1) = N_l^X 
     \frac{ \left\{(\vecl \odot \vecl_1) C_{l_1}^\cmb 
     + [\vecl \odot (\vecl-\vecl_1)] C_{|\vecl-\vecl_1|}^\cmb \right\} }
     {2C_{l_1}^\tot C_{|\vecl-\vecl_1|}^\tot},
\label{eqn:wx}
\end{equation}
\begin{equation}
     (N_l^X)^{-1} \equiv \int \frac{d^2\vecl_1}{(2\pi)^2}
     \frac{ \{
     (\vecl \odot \vecl_1) C^\cmb_{l_1} + [\vecl \odot
     (\vecl-\vecl_1)] C^\cmb_{|\vecl-\vecl_1|} \}^2}
     {2C_{l_1}^\tot C_{|\vecl-\vecl_1|}^\tot}.
\label{eqn:nl}
\end{equation}
Here, $C_{l}^\tot$ is the total temperature power spectrum and
can be written as a sum of the
lensed power spectrum, foregrounds, and detector noise: $C_{l}^\tot =
\widetilde C_l^\cmb + C_l^{\rm fore} + C_l^{\rm noise}$.  Filters similar to
these can be written down for the other quadratic combinations of
polarization and temperature. We do not write them out
explicitly here as they can be derived easily from published
expressions in the literature.  

In practice, one determines each Fourier mode $\phi(\vecl)$ [or
$\Omega(\vecl)$] by taking an appropriately weighted average of
all combinations of temperature (or polarization) Fourier modes
$\vecl_1$ and $\vecl_2$ that sum to $\vecl_1+\vecl_2=\vecl$. 
The only difference between the reconstruction of gradient
versus curl modes is whether to weight these combinations by
a dot product $\vecl_1 \cdot \vecl_2$ or by a curl $\vecl_1
\times \vecl_2$.  The quantity $N_l^{\rm X}$ is the noise, the
variance with which each Fourier mode $\phi(\vecl)$ or
$\Omega(\vecl)$ can be reconstructed.  Thus, when the $\phi$ or
$\Omega$ power spectrum is measured with these quadratic
estimators, the power spectra of the estimators will be,
\begin{equation}
\langle \hat{X}(\vecl) \hat{X}(\vecl')\rangle_{W^X} 
= (2\pi)^2 \delta(\vecl+\vecl') (C_l^{XX} + N_l^X) \, .
\label{spec}
\end{equation}
Of course, if the power spectra $C_l^{\rm t}$ and $C_l^\Theta$
are known, then the noise can be calculated independently and
subtracted to yield the desired $\phi$ or $\Omega$ power
spectra.

The orthogonality of the weightings in the filters for $\Omega$
and $\phi$ suggests that if we have a
deflection field that is a pure gradient, then the application
of the curl filter will give zero, and {\it vice versa}.
Although this is approximately correct, it is not precisely
true, as we now show.  Consider a deflection field that
is a pure gradient; i.e., it is described in terms of nonzero
$\phi(\vecl)$, with $\Omega=0$.  Suppose now that we measure
$\Omega$: taking Eq.~(\ref{eqn:quadest}) with $X=\Omega$, the only
possible source in the temperature field is due to the gradient, whereby 
\begin{widetext}
\begin{eqnarray}
     \langle \hat{\Omega}(\vecl) \rangle
     &=& \phi(\vecl)\int \frac{d^2\vecl_1}{(2\pi)^2} W^\Omega(\vecl,\vecl_1)
     \left \{(\vecl \cdot \vecl_1) C^\cmb_{l_1} + [\vecl \cdot
     (\vecl-\vecl_1)] C^\cmb_{|\vecl-\vecl_1|} \right\}
     \nonumber \\
     &=& \phi(\vecl) N_l^\Omega \int \frac{d^2\vecl_1}{(2\pi)^2} \frac{
     \left\{ (\vecl \times \vecl_1) C^\cmb_{l_1} + [\vecl \times
     (\vecl-\vecl_1)] C^\cmb_{|\vecl-\vecl_1|} \right\}}
     {2C_{l_1}^\tot C_{|\vecl-\vecl_1|}^\tot}
     \left\{ (\vecl \cdot \vecl_1) C^\cmb_{l_1} + [\vecl
     \cdot (\vecl-\vecl_1)] C^\cmb_{|\vecl-\vecl_1|} \right\} \, .
\label{eqn:orthogonal}
\end{eqnarray}
\end{widetext}
Despite the fact that the filter is designed to select out only the curl
contribution,  close inspection of this integral shows that it is not precisely
zero.  Note, however, that if $C_{l}^\Theta$  is a pure power-law --- thus, also
$C_{l}^{\rm t}$ is a power-law --- then the integral would vanish
identically.    In other words, the $\phi$ and $\Omega$ filters are orthogonal
only to the extent that $C_l$ behaves as  $\sim l^n$. The departure from
orthogonality is due to the presence of bumps and wiggles in the CMB anisotropy
power spectrum. The departure from a power-law spectrum also prevents the
construction of precisely orthogonal filters to separate the two modes exactly.

The integrand in Eq.~(\ref{eqn:orthogonal}) is nonzero
only for values of $l_1$ at which $C_{l_1}^\Theta$ and
$C_{|\vecl-\vecl_1|}^\Theta$  departs from the power-law.
The departure is not significant except 
when $l_1$ or $|\vecl-\vecl_1|$ enter the damping tail 
of the anisotropy spectrum. This contribution, however,
is suppressed by finite angular resolution of CMB experiments.  
We therefore expect that the integral in
Eq.~(\ref{eqn:orthogonal}) should be small, even if it is not
precisely zero.  A numerical evaluation confirms this argument;
we have found that the expression evaluates to well below
$10^{-10}\, \phi(\vecl)N_l^\Omega$ for $l$ values up to 5000.  We
therefore conclude that the reconstruction can be 
considered to be effectively orthogonal. 

This leads us to another point.  Cosmic shear, either through $\phi$ or
$\Omega$, leads to a correction to the observed CMB temperature that can be
written as $(\nabla\Theta)\cdot (\nabla \phi)$ or $(\nabla\Theta)\cdot (\nabla
\times \Omega)$. Suppose, however, that some other process lead to a correction
of the form $\Theta(\bn)f(\bn)$, where $f(\bn)$ is some function of position on
the sky.  For example, consider the Sunyaev-Zel'dovich effect \cite{SZ}.  The
thermal effect can be subtracted to a large extent through multifrequency
observations.  However, the kinetic-SZ effect has the same frequency dependence
as primordial fluctuations and is therefore indistinguishable. Suppose that
there is thus some unsubtracted SZ contribution to the measured CMB
fluctuation.  Then this will provide an angle-dependent multiplicative
correction to the primordial temperature.  Something similar (though not
precisely so) may occur through exotic phenomena such as primordial
non-Gaussianity \cite{Les04} or a spatially-varying fine-structure constant
\cite{kris}, for example. On the other hand, non-uniformities in the
instrumental gain may also mimic such an effect.  A quadratic estimator can be
constructed for $f(\bn)$, simply by removing the $\vecl$ vector dependences in
Eqs.~(\ref{eqn:wx}) and (\ref{eqn:nl}).  Again, the estimator for $f$ will be
close to orthogonal to those for $\phi$ and $\Omega$.  However, the
orthogonality will not be precise, and if there is a significant $f(\bn)$, then
it will show up in a reconstruction of $\phi$ and to a similar level in
$\Omega$.

\section{Cosmological Curl Sources}

\subsection{Primordial Gravitational Waves}
\label{sec3}

Our first example of a curl deflection is a background of
gravitational waves from inflation.
Refs.~\cite{KaiJaf97,stebbins} showed that gravitational waves
can act as gravitational lenses, and
Refs.~\cite{stebbins,Dodetal03} showed that lensing by
gravitational waves gives rise to a curl component in the
deflection angle.  Suppose there is a gravitational wave with
amplitude $h_{ij}$ (more precisely, the transverse traceless
tensor part of the metric perturbation), and suppose further
that we choose our line of sight to be (near the) ${\mathbf z}$
direction.  Then, $\Omega \propto \epsilon_{kl} \partial_k
h_{zl}$.  For example, if the gravitational wave propagates in
the ${\mathbf y}$ direction, then $\Omega \propto \partial_y
h_{zx}$ and the deflection is in the ${\mathbf \theta}_x$
direction with $\delta \theta_x \propto h_{zx}$.

Of course, the total deflection is an integral of all the
deflections along a line of sight, and for arbitrary line of
sight, the rotation is~\cite{Dodetal03},
\begin{eqnarray}
     \omega(\bn) &=&{1 \over 2}{1 \over \rad_s} \bn \cdot
     \left[\nabla_\bn \times \bfr (\bn, \rad_s)\right] \nonumber
     \\ 
     &=&-{1\over2}\int_0^{\rad_s} d\rad'
                \left[\bn \cdot (\nabla \times \bfH) \cdot \bn
                       \right]_{\left(\rad',\bn \rad'\right)} \, ,
\label{BornOmega}
\end{eqnarray}
where $\bfH$ is the transverse ($\nabla\cdot\bfH=0$), traceless
(Tr\,$\bfH=0$), tensor metric perturbation representing
gravitational waves. In the
above, $\rad$ is the radial distance from the observer and the
surface of last scattering is at $\rad_s$. The gravitational-wave amplitude
obeys a wave equation which, ignoring the presence of anisotropic stress
from neutrinos and other relativistic species at early times, takes the form
$\ddot{\bfH}-\nabla^2\bfH+2(\dot{a}/a)\bfH=0$, where the dot
denotes derivative with respect to conformal time.  We express the solution to
this equation  in the form of a transfer function, $T_{\rm(T)}(\bfk,\rad)$, 
whereby the Fourier amplitude evolves as  $H(\bfk,\rad) = \tilde H(\bfk)
T_{\rm(T)}(k,\rad)$ and $\tilde H(\bfk)$ is the initial amplitude of the wave.
In a purely dust-dominated universe (appropriate for the
long-wavelength modes of relevance here, which come into the
horizon at late time), $T_{\rm(T)}(k,\rad)=3j_1(k\rad)/(k\rad)$
(Ref.~\cite{pritchard} presents more precise expressions, but
they are not relevant for the calculation here).
Assuming isotropy, the three-dimensional spatial power spectrum of initial
metric fluctuations related to a stochastic background of gravitational waves
is 
\begin{equation}
\left\langle\tilde{H}_{(i)}(\bfk)\tilde{H}_{(j)}^*(\bfk')\right\rangle
=(2\pi)^3P_{\rm(T)}(k)\,\delta_{ij}\,\delta^{(3)}(\bfk-\bfk')\ ,
\end{equation}
where the two linear-polarization states of the gravitational wave are denoted
by $(i,j)$. In standard inflationary models, the primordial fluctuation spectrum
is predicted to be
\begin{equation}
P_{\rm(T)}(k) = A_T k^{n_T-3} \ .
\end{equation}
Inflationary models generally predict that $n_T \sim 0$ while the ratio of
tensor-to-scalar amplitudes, $r=A_T/A_S$,  is now constrained to be below 0.36
\cite{MelOdm03}. We will use the upper limit allowed when calculating the
inflationary-gravitational-wave (IGW) contribution.

Taking the spherical-harmonic moments of Eq.~(\ref{BornOmega}), 
the angular power spectrum of the rotational component is
\begin{eqnarray}
C_l^{\omega \omega} &=& {1\over2l+1}\sum_{m=-l}^l
\left\langle|\tilde{\omega}_{(l,m)}|^2\right\rangle \cr
&=&{2 \over \pi}\int
k^2\,dk\,P_{\rm(T)}(k)\,\left|T_l^\omega(k,\rad_s)\right|^2\, ,
\end{eqnarray}
where
\begin{eqnarray}
\label{LensingTransfer}
&&T_l^\omega(k,\rad)=\sqrt{{(l+2)!\over(l-2)!}}\,
  \int_0^\rad d\rad'\,T_{\rm(T)}(k,{\rad_s -\rad})\,
  {j_l(k \rad')\over{k \rad'^2}} \ . \nonumber \\
\end{eqnarray}
For comparison, the gradient components of the deflection angle involve the
projected density perturbations along the line of sight to the
last-scattering surface,
\begin{eqnarray}
\phi(\bm)&=&- 2 \int_0^{\rad_s} d\rad
\frac{(\rad_s-\rad)}{(\rad \rad_s)}
\Phi (\rad,\hat{{\bf m}}\rad ) \,,
\label{eqn:lenspotential}
\end{eqnarray}
where $\Phi$ is the potential associated with the large-scale mass
distribution. The angular power spectrum of these projected potentials are
\begin{eqnarray}
C_l^{\phi \phi} &=& {1\over2l+1}\sum_{m=-l}^l
\left\langle|\tilde{\phi}_{(l,m)}|^2\right\rangle \cr
&=&{2 \over \pi}\int k^2\, dk\,P_{\rm(\delta)}(k)\,\left|T_l^\phi(k,\rad_s)\right|^2 \, ,
\end{eqnarray}
where $P_{\rm(\delta)}(k)$ is the power spectrum of density perturbations,
including the transfer function, and
\begin{equation}
T_l^\phi(k,\rad_s) =-3\Omega_m \left(\frac{H_0}{k}\right)^2 \int_0^{\rad_s} d\rad' \frac{G(\rad')}{a(\rad')}
\frac{\rad_s-\rad'}{\rad' \rad_s} j_l(k \rad') \, ,
\end{equation}
with the growth of matter fluctuations given by $G(\rad)$, and $a(\rad)$ is the
scale factor. Here, we ignore the metric shear at the surface of last scattering
by the same background of waves.  As discussed in Ref.~\cite{Dodetal03}, the curl
component due to intervening deflections from the gravitational background is
small compared to the gradient component and the inclusion of metric shear only
leads to a further cancellation. Hence, our forthcoming proposal that the curl
component should be considered as a test of systematics will not
be affected adversely.

\begin{figure}[t]
\includegraphics[scale=0.4,angle=-90]{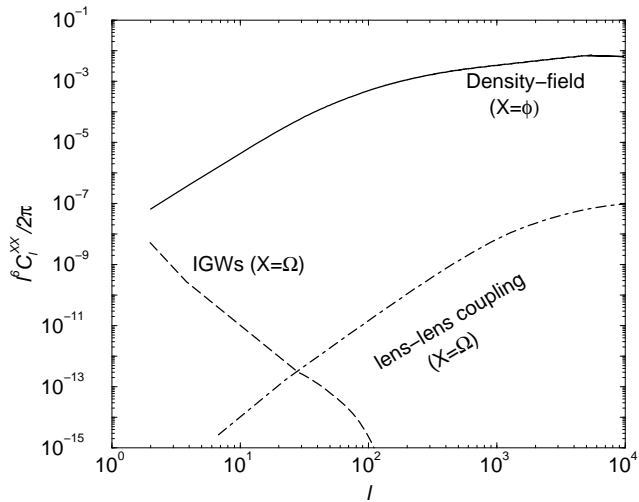}
\caption{Lensing-deflection power spectra. Here, we show the gradient
component from density perturbations (top curve), the curl
component from inflationary gravitational waves (dashed curve labeled `IGWs'),
and the curl component from second-order density perturbations.
(dot-dashed curve). We have taken the maximum IGW 
amplitude consistent with the current upper limit to
tensor-to-scalar ratio \cite{MelOdm03}.}
\label{spectra}
\end{figure}

\begin{figure*}[t]
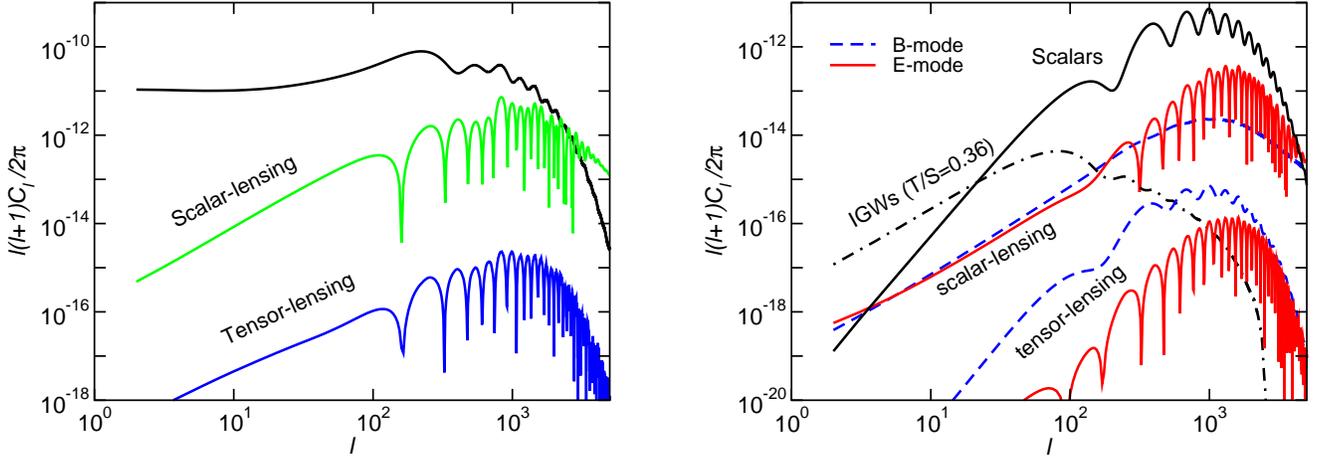

\includegraphics[scale=0.35,angle=0]{Fig2a.eps} \hspace{1.0cm}
\includegraphics[scale=0.35,angle=0]{Fig2b.eps}
\caption{The lensing modification to CMB power spectra for
density perturbations
and for gravitational waves. {\it Left:} Temperature
fluctuations.  The top curve is the primordial power spectrum.
The middle curve is the (additive) contribution to the
temperature power spectrum from lensing by density
perturbations, and the lower curve is for lensing by
gravitational waves, assuming the maximum IGW background
consistent with current CMB bounds.  Note that for the lower two curves, it is
$|\tilde C_l - C_l|$ which is plotted.  The negative
contribution of the $R$ term in Eq.~(\ref{eqn:ttflat}) allows
for $\tilde C_l-C_l$ to become negative.  Since the coherence
scale for $C_l^{\Omega\Omega}$ is so small, the lensed power
spectrum reflects closely at high $l$ the primordial temperature
power spectrum.  {\it Right:} Polarization.  These curves are:
(1) the top (solid) curve is the primordial E-mode power
spectrum; (2) the next-highest (solid) curve is the lensing correction to
the primordial E-mode power spectrum by density perturbations;
(3) the top dashed curve is the B-mode power spectrum resulting
from cosmic-shear conversion of E-modes by
density perturbations; (4) the lower dashed curve is the B-mode
power spectrum from lensing by foreground IGWs; and (5) the
lowest solid curve is the E-mode power
spectrum resulting from lensing by foreground IGWs.  The
dot-dash curve is the primordial B power spectrum from the
maximal IGW background allowed by current constraints.}
\label{cmb}
\end{figure*}

\subsection{Second-order density perturbations}

Gravitational lensing by density perturbations can give rise to
a curl component once we go to second order in the projected
potential $\phi$.  To see this, we first review the lowest-order
effect.  Suppose there is a lens at a distance $r_1$ along the
line of sight.  The deflection by this lens is $\delta\theta_i
\propto \partial_i \Phi_1$, where $\Phi_1$ is the gravitational
potential (not the projected potential) at $r_1$.  The
lowest-order deflection will therefore be written as the sum of
gradients perpendicular to the line of sight.  To second order
in $\Phi$, there can be deflection by two lenses at different
distances, $r_1$ and $r_2$, along the line of sight.  The
deflection by the first lens the ray encounters is $\propto
\partial_j \Phi_1$, and the deflection after encountering the
second lens is $\propto (\partial_i \partial_j
\Phi_2)(\partial_j \Phi_1)$ (this follows from the discussion,
e.g., in Section 3 of Ref.~\cite{CooHu02}).  To see that this
has nonvanishing curl, we take the curl: $\epsilon_{ik}
\partial_k [ (\partial_i \partial_j \Phi_2)(\partial_j
\Phi_1)] \propto \epsilon_{ik} (\partial_i \partial_j
\Phi_2)(\partial_j \partial_k \Phi_2)$ which does not generally
vanish.  

The full calculation of the curl power spectrum is then lengthy
but straightforward, and it is discussed in the context of
galaxy-based weak lensing surveys in Ref.~\cite{CooHu02}.  They
are explored in the context of CMB lensing in
Ref.~\cite{HirSel03}.  We do not repeat the derivation but refer
the reader to Ref.~\cite{CooHu02} for details.


\section{Calculations and Results}
\label{sec:results}

A comparison of the gravitational-wave and density-perturbation
(to second order) curl signals to the gradient
lensing signal is shown in Fig.~\ref{spectra}. 
[Note that
the anisotropy spectra for the lensing convergence and rotation
are related to the gradient and
curl by  $C_l^{\kappa \kappa}=(l^4/4) C_l^{\len \len}$ and $C_l^{\omega
\omega}=(l^4/4) C_l^{\Omega \Omega}$.] 
Here, we push the
gravitational-wave amplitude to the maximum allowed by current
data. A useful measure of the
relative importance of the two components is the rms deflection angle on the sky
given by $\theta^2_{\rm rms} = \int [d^2\vecl/(2\pi)^2] l^2
C_l^{XX}$. In the case of density perturbations (for $X=\phi$),
$\theta_{\rm rms}=7 \times 10^{-4}$
or roughly $2.5$ arcmins. The angular coherence scale, where the rms drops to
half its peak value, is about a degree. In the case of the strongest
gravitational wave background, the deflection angle is $\theta_{\rm rms}=7
\times 10^{-5}$ or 0.25 arcmins, but the angular coherence scale
is a few tens of degrees.  (Note that the $y$ axis in
Fig. \ref{spectra} is $l^6 C_l$, so the power spectra plotted
there are in fact very rapidly falling with $l$.  The rms
deflection angle is thus fixed primarily by the low $l$'s.)
In the case of second-order curl corrections, the coherence
scale is similar to that of density perturbations but
the amplitude is smaller by at least four to five orders of magnitude. The
resulting corrections to CMB anisotropies trace that of the density field, but
with a similar reduction in the overall amplitude. 

The effects on the CMB
anisotropy spectra are summarized in Fig.~\ref{cmb}. 
In the case of temperature, the
gravitational-wave-lensing correction is at least two orders of magnitude below
the temperature fluctuations associated with the  angular displacement
corrections due to the density field. We also summarize our results for the case
involving polarization anisotropies. In accord with Ref.~\cite{Dodetal03}, we
conclude that the correction resulting from the curl component is negligibly
small.

\begin{figure}[t]
\includegraphics[scale=0.4,angle=-90]{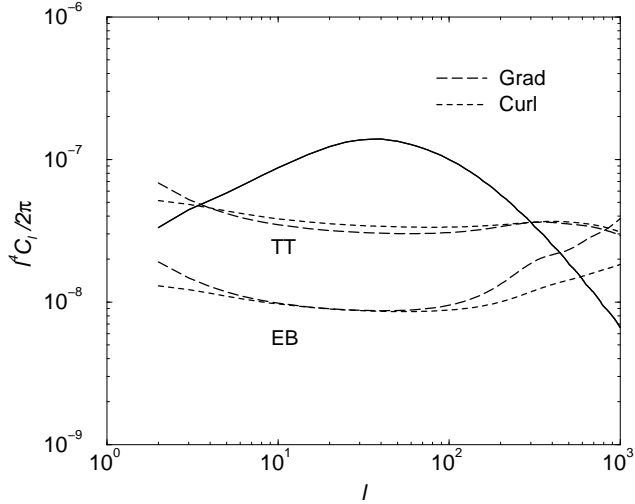}
\caption{Reconstructed unbinned noise spectra using the temperature-temperature
quadratic estimator and the EB polarization combination. We show noise for both
grad and curl modes. For reference, we also plot the power spectrum of
the deflection angle corresponding to the gradient component.}
\label{recon}
\end{figure}

We now turn to the reconstruction of the cosmic-shear pattern
with quadratic estimators for $\phi$ and $\Omega$.
Fig.~\ref{recon} shows the errors in the reconstruction for a
hypothetical
CMB experiment with a resolution of an arcminute and a
noise-equivalent temperature of 1~$\mu$K sec$^{1/2}$ over one
year of integration. We show the
reconstruction for both temperature maps (top lines), and for the EB quadratic
combination which was shown in Ref.~\cite{HuOka02} to be the
best combination to
extract lensing information from CMB data. Note that one generally reconstructs
the gradient and curl components of the deflection field with roughly the same
signal-to-noise ratio. The gradient component, however, dominates since it is
sourced by the large-scale mass distribution, while the curl component
is subdominant given that the amplitude of the tensor
contribution to the CMB quadrupole is limited by current CMB
data to be less than $30\%$ of that due to scalar perturbations.

\section{The Curl as a Diagnostic}
\label{sec:diagnostic}

Is it useful to reconstruct the curl component as there is virtually no signal?
Here, we suggest that a reconstruction may be useful to
identify non-Gaussian signals, both due to primary
effects, such as  primordial non-Gaussianity \cite{Les04} or
perhaps variable fine-structure constant \cite{kris}, and
secondary anisotropies.  As discussed above, although the
thermal SZ effect can can be removed largely
from multifrequency data \cite{cooray}, the kinetic-SZ effect cannot.
Such an unsubtracted secondary contribution to CMB fluctuations
will give rise to an additional noise-bias term in
Eq.~(\ref{spec}), whereby
\begin{equation}
     \langle \hat{X}(\vecl) \hat{X}(\vecl')\rangle
     = (2\pi)^2 \delta(\vecl+\vecl') (C_l^{XX} + N_l^X + S_l^X) \, .
\end{equation}
When secondary non-Gaussianities are not properly accounted for, the additional
noise-bias term takes the form 
\begin{eqnarray}
      && S_l^X = \left(N_l^X\right)^2 \nonumber \\
      &\times& \int \frac{d^2\vecl_1}{(2\pi)^2} \int
      \frac{d^2\vecl_2}{(2\pi)^2} 
      \frac{ \left\{ (\vecl \odot \vecl_1) C^\cmb_{l_1} + [\vecl \odot
      (\vecl-\vecl_1)] C^\cmb_{|\vecl-\vecl_1|} \right\} }
      {2C_{l_1}^\tot C_{|\vecl-\vecl_1|}^\tot} \nonumber \\
      &\times& \frac{ \left\{ (\vecl \odot \vecl_2) C^\cmb_{l_2}
      + [\vecl \odot (\vecl-\vecl_2)] C^\cmb_{|\vecl-\vecl_2|}
      \right\} }{2C_{l_2}^\tot C_{|\vecl-\vecl_2|}^\tot} \nonumber \\
      &\times&\langle \Theta^s(\vecl_1) \Theta^s(\vecl-\vecl_1)
      \Theta^s(\vecl_2) \Theta^s(-\vecl-\vecl_2) \rangle \, ,
\end{eqnarray}
where $\langle \Theta^s(\vecl_1) \Theta^s(\vecl-\vecl_1) \Theta^s(\vecl_2)
\Theta^s(-\vecl-\vecl_2) \rangle$ is the four-point correlator of the
contaminant foreground or primordial non-Gaussianity with its
anisotropy written in Fourier-space as $\Theta^s(\vecl)$. This
correlator can be decomposed as
\begin{eqnarray}
     && \langle \Theta^s(\vecl_1) \Theta^s(\vecl-\vecl_1)
     \Theta^s(\vecl_2) \Theta^s(-\vecl-\vecl_2) \rangle
     \nonumber \\
     && = 
     2C^{ss}_{l_1} C^{ss}_{|\vecl-\vecl_1|} \delta(\vecl_1+\vecl_2) 
     \nonumber \\  
     &&  +  T^s(\vecl_1,\vecl-\vecl_1,\vecl_2,-\vecl-\vecl_2). 
\end{eqnarray}
The Gaussian piece leads to a noise bias 
\begin{eqnarray}
S_l^X &=& \left(N_l^X\right)^2 \nonumber \\
     &=& \times \int \frac{d^2\vecl_1}{(2\pi)^2} \frac{ \left\{ (\vecl
     \odot \vecl_1) C^\cmb_{l_1} + [\vecl \odot (\vecl-\vecl_1)]
     C^\cmb_{|\vecl-\vecl_1|}\right\}^2}
     {\left[2C_{l_1}^\tot C_{|\vecl-\vecl_1|}^\tot\right]^2}
     \nonumber \\
     \, \, & \times & 2\, C^{ss}_{l_1} C^{ss}_{|\vecl-\vecl_1|} \, ,
\label{eqn:Gau}
\end{eqnarray}
which can be absorbed into $N_l^X$ with a proper definition of
the normalization
factor, and where $C_l^{\rm tot}$ also include foregrounds and secondary power
spectra. However, the non-Gaussian nature of the foreground cannot be ignored
and this results in a bias that cannot be removed by a renormalization. This
noise is,
\begin{eqnarray}
     S_l^X &=& \left(N_l^X\right)^2 \int \frac{d^2\vecl_1}{(2\pi)^2} \int
     \frac{d^2\vecl_2}{(2\pi)^2}  \nonumber \\
     &\times &\frac{ \left\{ (\vecl \odot \vecl_1) C^\cmb_{l_1} + [\vecl \odot
     (\vecl-\vecl_1)] C^\cmb_{|\vecl-\vecl_1|} \right\}}
     {2C_{l_1}^\tot C_{|\vecl-\vecl_1|}^\tot} \nonumber \\
     &\times& \frac{ \left\{ (\vecl \odot \vecl_2) C^\cmb_{l_2} + [\vecl
     \odot (\vecl-\vecl_2)] C^\cmb_{|\vecl-\vecl_2|} \right\}}
     {2C_{l_2}^\tot C_{|\vecl-\vecl_2|}^\tot} \nonumber \\
     &\times& T^s(\vecl_1,\vecl-\vecl_1,\vecl_2,-\vecl-\vecl_2) \, .
\end{eqnarray}
The angular dependence of the foreground trispectrum is important: if the
trispectrum were to depend on the length of the vectors alone, the averaging
would result in significant suppression of this noise bias. 

In the presence of an additional secondary trispectrum, for lensing
reconstruction,
\begin{equation}
     \langle \hat{\phi}(\vecl) \hat{\phi}(\vecl')\rangle
     = (2\pi)^2 \delta(\vecl+\vecl') (C_l^{\phi \phi} + N_l^\phi
     +S_l^\phi) \, .
\end{equation}
While $N_l^\phi$ can be established based on noise properties, one cannot
separate the signal $C_l^{\phi \phi}$ of interest from the confusion
$S_l^\phi$. Such a situation has already been observed, for example, in
numerical simulations of the CMB lensing reconstruction process where in the
presence of a kinetic-SZ component an additional noise bias was suggested
\cite{Amb04}. The presence of such a noise bias was readily detectable given
that the input mass spectrum, or alternatively the cosmology,
was known {\it a priori}. In the real case, one is interested in
measuring these quantities from
the mass spectrum determined from CMB lensing. Thus, the presence of a noise bias
cannot easily be established since the bias is degenerate with other unknowns.

The curl component, however, provides a useful method to establish the presence
of such a noise bias which can be used to correct the gradient component or to
allow for an accounting of the bias when cosmological parameters are measured.
This follows from the fact that all signals of interest in the curl component are
negligibly small  such that the resulting reconstruction only leads to 
\begin{eqnarray}
     \langle \hat{\Omega}(\vecl) \hat{\Omega}(\vecl')\rangle &=&
     (2\pi)^2 \delta(\vecl+\vecl') (N_l^\Omega +S_l^\Omega) \, .
\end{eqnarray}
Since $N_l^\Omega$, like $N_l^\phi$ can be predicted from the
measured CMB power spectra,
any excess noise in the reconstruction will suggest either systematics or the
presence of an additional non-Gaussian signal that is contributing via
$S_l^\Omega$. Even though the excess noises in the gradient and curl of the
deflection field are different---i.e., $S_l^\Omega$ vs
$S_l^\phi$---the origin of the excess noise could very well be
the same with the only differences resulting
from  variations in the two filters for the two modes. In general, any detection
of excess noise in the curl component should suggest a bias in the gradient
component.  Since filter shapes are known {\it a priori}, 
one should be able to establish some estimate on the expected excess
noise in the gradient, given the excess noise in the curl.
If this excess noise is significant, then the
dominance of a systematic effect in the reconstruction is clearly established.
Currently, there is no mechanism to either estimate or establish
the presence of
a systematic noise component in the CMB lensing analysis. Thus, we suggest that
the curl component be used as a monitor of systematic effects and to understand
if the  $C_l^{\phi \phi}$ reconstruction is affected through $S_l$ by
non-Gaussian secondary effects  and foregrounds.

In general, we do not expect effects such as primordial non-Gaussianity
\cite{Les04} to be a significant concern for lensing reconstruction of the
deflection-potential statistics. Given the noise levels to the
reconstruction, as
shown in Fig.~\ref{recon}, one can establish the minimum amplitude for which
systematic effects or additional noise biases, as described by $S_l$, can be
detected via
\begin{equation}
\sigma_A^{-2} = \sum_l \frac{1}{\sigma_l^2}
\left(\frac{\partial S_l}{\partial A}\right)^2  \, .
\end{equation}
Here $\sigma_l = \sqrt{2/(2l+1)} N_l$, under the assumption of no signal in the
curl component. Using the estimated noise levels, from the EB combination of
polarization maps to reconstruct the curl component, for example, one can
establish systematic effects down to a level of 0.1\% from the amplitude of the
potential-fluctuation power spectrum.  In the cosmic-shear
simulations of Ref.~\cite{Amb04}, noise biases at the level of 30\% or
more were found.  We surmise that some of this may be due to
conversion of kinetic-SZ corrections to a deflection angle,
although this probably does not account for all the excess noise.
A study of the curl component may help clarify
the nature of such noise biases in the simulation.

\section{Summary}
\label{sec5}


Lensing by gravitational waves is expected to give rise to a
cosmic-shear pattern where the deflection angle has a curl
(or transverse-vector) component, as opposed to the curl-free
pattern expected by cosmic shear by density perturbations (to
linear order in the perturbation amplitude).  To second order in
the perturbation amplitude, lensing can also give rise to a curl
component.  
For a primordial background of gravitational waves from inflation with a
normalization given by the current upper limit to the
tensor-to-scalar ratio, the
corrections to the CMB power spectra are generally two orders of
magnitude below those of the cosmic shear due to density
perturbations, and the curl component from higher-order lensing
effects is also small. 
The curl component can be reconstructed with quadratic
estimators analogous to those developed to measure the gradient component.
Given the small signal expected in the curl, this
component can potentially be used as a probe of systematic
effects and foregrounds for next-generation CMB experiments

\begin{acknowledgments}
MK thanks C.~Vale and C.~Hirata for useful discussions.  This
work was supported in part by NASA NAG5-11985 and DoE
DE-FG03-92-ER40701 at Caltech and NSF AST-0349213 at Dartmouth.  
\end{acknowledgments}


\end{document}